\documentclass[aps,preprint]{revtex4}%
\usepackage{amsfonts}
\usepackage{amsmath}
\usepackage{amssymb}
\usepackage{graphicx}%
\providecommand{\U}[1]{\protect\rule{.1in}{.1in}}

\begin{document}

\title{One- and two-dimensional modes in the complex Ginzburg-Landau
equation with a trapping potential}
\author{Thawatchai Mayteevarunyoo$^{1}$, Boris A. Malomed$^{2,3}$, and
Dmitry V. Skryabin$^{3,4}$}

\affiliation{$^{1}$Department of Electrical and Computer Engineering, Faculty of
Engineering, Naresuan University, Phitsanulok 65000, Thailand\\
$^{2}$Department of Physical Electronics, School of Electrical Engineering,
Faculty of Engineering, and Center for Light-Matter Interaction, Tel Aviv
University, Tel Aviv 69978, Israel\\
$^{3}$ITMO University, St. Petersburg 197101, Russia\\
$^{4}$Department of Physics, University of Bath, Bath, BA2 7AY, UK}


\begin{abstract}
We propose a new mechanism for stabilization of confined modes in lasers and
semiconductor microcavities holding exciton-polariton condensates, with
spatially uniform linear gain, cubic loss, and cubic self-focusing or
defocusing nonlinearity. We demonstrated that the commonly known background
instability driven by the linear gain can be suppressed by a combination of
a harmonic-oscillator trapping potential and effective diffusion. Systematic
numerical analysis of one- and two-dimensional (1D and 2D) versions of the
model reveals a variety of stable modes, including stationary ones,
breathers, and quasi-regular patterns filling the trapping area in the 1D
case. In 2D, the analysis produces stationary modes, breathers, axisymmetric
and rotating crescent-shaped vortices, stably rotating complexes built of up
to $8$ individual vortices, and, in addition, patterns featuring vortex
turbulence. Existence boundaries for both 1D and 2D stationary modes are
found in an exact analytical form, and an analytical approximation is
developed for the full stationary states.

\end{abstract}
\maketitle

\section{Introduction}

Formation of optical patterns in nonlinear media, that include pump, gain
and loss mechanisms, such as laser cavities, as well as in many other
physical settings, is frequently modeled by complex Ginzburg-Landau
equations (CGLEs) \cite{Hohenberg}-\cite{Rosanov}. In many cases, modes of
primary interest are localized ones, supported by a stable balance between
the gain and loss, as well as between the self-focusing nonlinearity and
transverse diffraction, thus forming dissipative solitons \cite{Akhmediev}-%
\cite{Rosanov2}. In particular, the simplest one-dimensional (1D) cubic CGLE
gives rise to exact localized solutions in the form of \textrm{sech}
solitons with the phase chirp \cite{Stewartson,Stenflo}. However, a
well-known problem for this simplest solution is that it is unstable,
because the spatially uniform linear gain renders the zero background of the
solitons unstable. One possibility for stabilization of dissipative solitons
is offered by the adoption of the cubic-quintic nonlinearity, in which case
the linear gain is replaced by linear loss, making the background stable,
while the gain is supplied by the cubic term, and the quintic dissipation
provides for the overall stabilization of the model \cite{Sergeev}. In
optics, this model can be realized using a combination of a usual linear
amplifier with a saturable absorber \cite{Leblond,Wise}. In particular, the
CGLE with the cubic-quintic nonlinearity readily creates stable fundamental
2D dissipative solitons and their generalizations with the intrinsic
vorticity \cite{Lucian}. These solitons can be additionally stabilized by
the addition of the harmonic-oscillator (HO) trapping potential to the
model, which also helps to build stable rotating two- and three-vortex
complexes trapped in this potential \cite{Skarka}.

On the other hand, keeping the nonlinearity in the basic cubic form, stable
1D dissipative solitons were predicted in two-component systems, where the
additional lossy component, linearly-coupled to the gain-carrying one,
provides for the stabilization of the background \cite{Winful}-\cite{Willie}%
. In particular, assuming that the nonlinearity is present only in the
active component, it is possible to find 1D dissipative solitons
analytically, in the form of the chirped \textrm{sech} ansatz, which have a
considerable stability domain \cite{Atai}. In single-component settings with
the cubic-only nonlinearity, stability of localized modes can also be
secured if the linear gain is supplied in a confined area embedded inside a
linear-loss background \cite{Kartashov,JOSAB}.

Another approach to the stabilization of 1D dissipative solitons in the
cubic medium was elaborated in Ref. \cite{Barcelona}, also using spatial
inhomogeneity, but in a different form: the instability of the background
induced by the linear spatially uniform gain can be suppressed if the local
coefficient of the cubic losses grows from the center to periphery faster
than $|x|$. While this setting seems somewhat \textquotedblleft exotic", it
is interesting to test a possibility to build a more straightforward one, by
combining the uniform linear gain and the HO trapping potential, $\sim x^{2}$%
, which may approximate deep guiding channels in various photonic settings
\cite{Agrawal,Konotop}. Indeed, although the linear gain amplifies
perturbations far from the localized mode, the HO potential will push the
perturbations towards the core area, where the localized mode may have a
chance to suppress them with the help of the usual cubic loss. Recently,
this possibility was tested in the two-dimensional (2D)\ model of a spinor
(two-component) exciton-polariton condensate, with its components linearly
mixed by the spin-orbit coupling (SOC) \cite{Sakaguchi}. It was found that
the interplay of the linear gain, HO trap, complex cubic nonlinearity, and
SOC gives rise to families of stable 2D dissipative solitons in the form of
mixed modes (so called because they combine zero-vorticity and vortex terms
in each component), vortex-antivortex complexes, and semi-vortices (which
feature vorticity in one component, provided that the Zeeman splitting
between the components is present). However, the stability is only possible
if the model also includes an imaginary part ($\eta $)\ of the diffraction
coefficient, i.e., spatial filtering \cite{Rosanov3}. It may represent
ionization of the medium \cite{ionization}, or relatively poor finesse of
the laser cavity \cite{filter1}-\cite{filter3}, or diffusion of excitons in
semiconductor microcavities \cite{diffusion1}-\cite{diffusion5}). The
stability of the above-mentioned 2D confined modes is secured if $\eta $
exceeds a certain threshold value.

These results suggest a possibility to consider a setting in a still more
fundamental form of the single-component cubic CGLE with the linear gain,
spatial filtering (effective diffusion), and the HO potential, in 1D and 2D
geometries alike. The model may apply both to the optical laser cavities and
microcavities populated by the exciton-polariton condensate. This analysis
is the objective of the present work and it will be chiefly performed by
means of numerical methods. Nevertheless, the existence boundary for
dissipative solitons is obtained in an \emph{exact analytical form}, in the
1D and 2D cases alike, due to the fact that the linearized form of both the
1D and 2D CGLE, including the spatial filtering and HO potential, admits
(novel) exact solutions.

The paper is organized as follows. The models, both 1D and 2D ones, are
introduced in Section II, where we also produce exact analytical solutions
of the linearized CGLEs, which determine the existence boundaries of the 1D
and 2D confined states, and some additional approximate analytical results
for nonlinear stationary modes. Numerical results for the 1D and 2D models
are reported, severally, in Sections III and IV. It is found that the
parameter space of the 1D CGLE is populated by stable stationary modes or
breathers (the latter occupying a narrow stripe) and persistent multi-peak
quasi-regular states, which fill out the entire spatial domain confined by
the HO potential. The 2D model also gives rise to stable stationary modes
and breathers, as well as to vortices (both isotropic ones and deformed
rotating \textquotedblleft crescents") and rotating robust multi-vortex
bound states, with the net vorticity up to $S=8$. The parameter space of the
2D model also features an area of vortex turbulence. The paper is concluded
by Section V.

\section{The model and analytical results}

\subsection{Complex Ginzburg-Landau equations}

All the ingredients of the setting under the consideration, which was
outlined in the introduction, are included in the 1D CGLE, with spatial
coordinate $x$, which governs the evolution of local amplitude $\psi $ of
the electromagnetic wave:%
\begin{gather}
i\frac{\partial \psi }{\partial t}=-\frac{1}{2}(1-i\eta )\frac{\partial
^{2}\psi }{\partial x^{2}}  \notag \\
-\sigma |\psi |^{2}\psi +i(\gamma -\Gamma |\psi |^{2})\psi +\frac{1}{2}%
\Omega ^{2}x^{2}\psi .  \label{1}
\end{gather}%
Here $\eta \geq 0$ is the above-mentioned effective diffusivity (in
particular, for excitons in a semiconductor microcavity), $\gamma >0$ and $%
\Gamma >0$ are, respectively, the linear gain and cubic loss, the
nonlinearity coefficient takes one of three normalized values,%
\begin{equation}
\sigma =+1,0,-1,  \label{sigma}
\end{equation}%
with $+1$ and $-1$ corresponding, severally, to the self-focusing and
defocusing, and coefficient $\Omega ^{2}$ determines the strength of the
trapping HO potential. Variable $t$ in Eq. (\ref{1}) is the temporal one in
the model of the exciton-polariton condensate, or propagation coordinate in
the model of the planar laser cavity in optics. On the other hand, a
confined 1D laser cavity may be described by Eq. (\ref{1}) with $t$
remaining the temporal variable, similar to models of the Lugiato-Lefever
type \cite{LL}. It is relevant to mention that more fundamental models of
laser cavities are based on coupled equations of the Maxwell-Bloch type;
however, in many case, the atomic degrees of freedom may be adiabatically
eliminated, reducing the model to the field-evolution equation of the CGLE
type \cite{NM}.

It is also relevant to mention that, depending on a particular setting, the
instability of the zero state in laser cavities may emerge at a finite
wavenumber, rather than at the infinitely small one (the short-wave
instability, instead of its long-wave counterpart). In that case, the basic
nonlinear model is represented not by the CGLE, but by the complex
Swift-Hohenberg equation \cite{LMN}. The latter model may be a subject for a
separate work.

Keeping normalization condition (\ref{sigma}), one can still perform
rescaling of Eq. (\ref{1}), so as to make coefficients $\gamma $ and $\Gamma
$ equal, therefore we also fix relation%
\begin{equation}
\gamma =\Gamma .  \label{gg}
\end{equation}%
Thus, Eq. (\ref{1}), subject to conditions (\ref{sigma}) and (\ref{gg}), has
three free coefficients: $\eta $, $\gamma =\Gamma $, and $\Omega ^{2}$. The
sign coefficient (\ref{sigma}) should be added to this set of control
parameters.

Stationary states in the polariton condensate with real chemical potential $%
\mu $ (alias propagation constant, $-\mu $, in the models of planar laser
cavities, with $t$ playing the role of the propagation coordinate) are
looked for as solutions to Eq. (\ref{1}) in the form of%
\begin{equation}
\psi \left( x,t\right) =e^{-i\mu t}u(x),  \label{psiu}
\end{equation}%
with complex function $u(x)$ satisfying the equation%
\begin{gather}
\mu u=-\frac{1}{2}(1-i\eta )\frac{d^{2}u}{dx^{2}}  \notag \\
-\sigma |u|^{2}u+i(\gamma -\Gamma |u|^{2})u+\frac{1}{2}\Omega ^{2}x^{2}u.
\label{u}
\end{gather}%
Stationary states are characterized by the integral power (alias the norm),%
\begin{equation}
N=\int_{-\infty }^{+\infty }\left\vert u(x)\right\vert ^{2}dx,  \label{N}
\end{equation}%
even if its not a dynamical invariant of the present dissipative model.

The stability analysis for stationary states (\ref{psiu}) is based on
introducing a perturbed solution,%
\begin{equation}
\psi (x,t)=e^{-i\mu t}\left\{ u(x)+\varepsilon \left[ v(x)\exp \left(
\lambda t\right) +w^{\ast }(x)\exp \left( \lambda ^{\ast }t\right) \right]
\right\} ,  \label{eps}
\end{equation}%
where $\varepsilon $ determines the smallness of the perturbation, $v(x)$
and $w(x)$ are complex components of the perturbation eigenmode, and complex
$\lambda $ is the respective stability eigenvalue. The substitution of
ansatz (\ref{eps}) in Eq. (\ref{1}) and linearization with respect to $%
\varepsilon $ leads to the system of stationary linear equations:%
\begin{gather}
\left( \mu +i\lambda -i\gamma \right) v=-\frac{1}{2}(1-i\eta )\frac{d^{2}v}{%
dx^{2}}+\frac{1}{2}\Omega ^{2}x^{2}v  \notag \\
-2\left( \sigma +i\Gamma \right) |u(x)|^{2}v-\left( \sigma +i\Gamma \right)
\left( u(x)\right) ^{2}w,  \notag \\
\left( \mu -i\lambda +i\gamma \right) w=-\frac{1}{2}(1+i\eta )\frac{d^{2}w}{%
dx^{2}}+\frac{1}{2}\Omega ^{2}x^{2}w  \notag \\
-2\left( \sigma -i\Gamma \right) |u(x)|^{2}w-\left( \sigma -i\Gamma \right)
\left( u^{\ast }(x)\right) ^{2}v.  \label{pert}
\end{gather}%
Instability is accounted for by eigenvalues with $\mathrm{Re}(\lambda )>0$.

The 2D model is based on the corresponding version of Eq. (\ref{1}), which
is written, in terms of the polar coordinates, $\left( r,\theta \right) $,
as
\begin{gather}
i\frac{\partial \psi }{\partial t}=-\frac{1}{2}(1-i\eta )\left( \frac{%
\partial ^{2}}{\partial r^{2}}+\frac{1}{r}\frac{\partial }{\partial r}+\frac{%
1}{r^{2}}\frac{\partial ^{2}}{\partial \theta ^{2}}\right) \psi  \notag \\
-\sigma |\psi |^{2}\psi +i(\gamma -\Gamma |\psi |^{2})\psi +\frac{1}{2}%
\Omega r^{2}\psi ,  \label{2D}
\end{gather}%
with coefficients subject to conditions (\ref{sigma}) and (\ref{gg}). Its
stationary solutions with integer vorticity $S=0,1,2,...$ are looked for as
\begin{equation}
\psi \left( x,y,t\right) =e^{-i\mu t+iS\theta }u(r),  \label{vortex}
\end{equation}%
where complex function $u(r)$ satisfies the radial equation:
\begin{gather}
\mu u=-\frac{1}{2}(1-i\eta )\left( \frac{d^{2}}{dr^{2}}+\frac{1}{r}\frac{d}{%
dr}-\frac{S^{2}}{r^{2}}\right) u  \notag \\
-\sigma |u|^{2}u+i(\gamma -\Gamma |u|^{2})u+\frac{1}{2}\Omega r^{2}u,
\label{u2D}
\end{gather}%
with boundary condition $u(r)\sim r^{S}$ at $r\rightarrow 0$. The integral
power (norm) of the 2D state is%
\begin{equation}
N=2\pi \int_{0}^{\infty }\left\vert u(r)\right\vert ^{2}rdr.  \label{N2D}
\end{equation}

For the stability test, perturbed 2D solutions were looked for as [cf. Eq. (%
\ref{eps})]%
\begin{gather}
\psi (x,t)=\left\{ e^{-i\mu t}u(r)+\varepsilon \left[ v(xr)\exp \left(
\lambda t+im\theta \right) \right. \right.  \notag \\
\left. \left. +w^{\ast }(r)\exp \left( \lambda ^{\ast }t-im\theta \right)
\right] \right\} ,  \label{eps2D}
\end{gather}%
where integer $m$ is an independent angular index of the perturbation mode,
and the radial equations for perturbation amplitudes are%
\begin{gather}
\left( \mu +i\lambda -i\gamma \right) v=-\frac{1}{2}(1-i\eta )\left[ \frac{%
d^{2}}{dr^{2}}+\frac{1}{r}\frac{d}{dr}-\frac{\left( S+m\right) ^{2}}{r^{2}}%
\right] v  \notag \\
+\frac{1}{2}\Omega ^{2}r^{2}v-2\left( \sigma +i\Gamma \right)
|u(x)|^{2}v-\left( \sigma +i\Gamma \right) \left( u(x)\right) ^{2}w,  \notag
\\
\left( \mu -i\lambda +i\gamma \right) w=-\frac{1}{2}(1+i\eta )\left[ \frac{%
d^{2}}{dr^{2}}+\frac{1}{r}\frac{d}{dr}-\frac{\left( S-m\right) ^{2}}{r^{2}}%
\right] w  \notag \\
+\frac{1}{2}\Omega ^{2}r^{2}w-2\left( \sigma -i\Gamma \right)
|u(x)|^{2}w-\left( \sigma -i\Gamma \right) \left( u^{\ast }(x)\right) ^{2}v.
\label{pert2D}
\end{gather}%
This eigenvalue problems defined by Eqs. (\ref{pert}) and (\ref{pert2D})
were solved by means of numerical methods.

\subsection{Analytical solutions}

The amplitude of stationary states vanishes at their existence boundary,
where, accordingly, one-dimensional equation (\ref{u}) is replaced by its
linearized version,
\begin{equation}
\mu u=-\frac{1}{2}(1-i\eta )\frac{d^{2}u}{dx^{2}}+i\gamma u+\frac{1}{2}%
\Omega ^{2}x^{2}u.  \label{linear}
\end{equation}%
Straightforward consideration of Eq. (\ref{linear}) demonstrates that it
gives rise to an\emph{\ exact} ground-state (GS) solution, with arbitrary
amplitude $A_{0}$,%
\begin{equation}
u_{\mathrm{lin}}^{(\mathrm{1D})}(x)=A_{0}\exp \left( -\frac{\Omega }{2\sqrt{%
1-i\eta }}x^{2}\right),  \label{lin}
\end{equation}%
\begin{equation}
\mu _{\mathrm{lin}}^{(\mathrm{1D})}=\frac{\Omega \eta }{2\sqrt{2\left( \sqrt{%
1+\eta ^{2}}-1\right) }}.  \label{mu_thr}
\end{equation}%
provided that diffusivity $\eta $\ is related to linear gain $\gamma $ as
follows:%
\begin{equation}
\gamma _{\mathrm{thr}}^{(\mathrm{1D})}=\frac{\Omega }{2\sqrt{2}}\sqrt{\sqrt{%
1+\eta ^{2}}-1}.  \label{eta_thr-exact}
\end{equation}%
This exact solution follows the commonly known pattern of the GS wave
function of the one-dimensional HO potential in quantum mechanics. However,
the complex coefficient in front of the second derivative in Eq. (\ref%
{linear}) introduces an essential difference, in the form of the \textit{%
spatial chirp }of the wave function (a phase term $\sim x^{2}$).

Equation (\ref{eta_thr-exact}) determines the existence threshold of the GSs
of the full 1D nonlinear model, which are indeed found, in the numerical
form, precisely at $\gamma >\gamma _{\mathrm{thr}}^{(\mathrm{1D})}$, as
shown below. In the absence of the nonlinearity ($\sigma =\Gamma =0$), Eq. (%
\ref{1}) gives rise to an exponentially growing solution at $\gamma >\gamma
_{\mathrm{thr}}^{(\mathrm{1D})}$:%
\begin{gather}
\psi _{\mathrm{lin}}^{(\mathrm{1D})}(x)=A_{0}\exp \left[ \left( \gamma
-\gamma _{\mathrm{thr}}^{(\mathrm{1D})}\right) t\right]  \notag \\
\times \exp \left( -\frac{\Omega }{2\sqrt{1-i\eta }}x^{2}\right) .
\label{exp}
\end{gather}

Similarly, the linearized version of the 2D stationary equation (\ref{u2D}),
\begin{gather}
\mu u=-\frac{1}{2}(1-i\eta )\left( \frac{d^{2}}{dr^{2}}+\frac{1}{r}\frac{d}{%
dr}-\frac{S^{2}}{r^{2}}\right) u  \notag \\
+i\gamma u+\frac{1}{2}\Omega r^{2}u,  \label{linear2D}
\end{gather}%
gives rise to exact solutions for both the GS ($S=0$) and vortices ($S\geq 1$%
):%
\begin{equation}
u_{\mathrm{lin}}^{(\mathrm{2D})}(r)=A_{0}r^{S}\exp \left( -\frac{\Omega }{2%
\sqrt{1-i\eta }}r^{2}\right) ,  \label{lin2D}
\end{equation}%
\begin{equation}
\mu _{\mathrm{lin}}^{(\mathrm{2D})}=\frac{(1+S)\Omega \eta }{\sqrt{2\left(
\sqrt{1+\eta ^{2}}-1\right) }},  \label{mu2D}
\end{equation}%
the corresponding threshold being
\begin{equation}
\gamma _{\mathrm{thr}}^{(\mathrm{2D})}=\left( 1+S\right) \Omega \sqrt{\frac{1%
}{2}\left( \sqrt{1+\eta ^{2}}-1\right) }.  \label{eta_thr_2D}
\end{equation}%
These exact solutions follow the pattern of the GS and eigenstates carrying
the orbital angular momentum for the two-dimensional HO potential, the
difference from the quantum-mechanical wave functions being the presence of
the \textit{radial chirp}, i.e., a phase term $\sim r^{2}$.

At $\gamma >\gamma _{\mathrm{thr}}^{(\mathrm{2D})}$, the parameter space of
the 2D nonlinear model is populated by several types of robust static and
dynamical modes, as shown below, while, in the absence of the nonlinearity ($%
\sigma =\Gamma =0$), an exponentially growing solution to Eq. (\ref{2D}) is
[cf. Eq. (\ref{exp})]%
\begin{gather*}
\psi _{\mathrm{lin}}^{(\mathrm{2D})}(r)=A_{0}\exp \left[ \left( \gamma
-\gamma _{\mathrm{thr}}^{(\mathrm{2D})}\right) t\right] \\
\times r^{S}\exp \left( -\frac{\Omega }{2\sqrt{1-i\eta }}r^{2}\right) .
\end{gather*}

For the 1D nonlinear model (\ref{1}), an approximate analytical solution can
be found in the case of $\sigma =0$, $\Gamma >0$ (i.e., if the nonlinearity
is represented solely by the cubic loss), treating the gain and loss terms
as small perturbations. In the zero-order approximation, the stationary
solution is represented by the GS of the HO potential, $u_{0}(x)=A_{0}^{%
\mathrm{(1D)}}\exp \left( -\Omega x^{2}/2\right) ,~\mu =\Omega /2$ [cf. Eq. (%
\ref{lin})], where amplitude $A_{0}$ is determined, in the first-order
approximation, by the power-balance condition:%
\begin{equation}
\int_{-\infty }^{+\infty }\left( \gamma |u|^{2}-\Gamma |u|^{4}-\frac{1}{2}%
\eta \left\vert \frac{du}{dx}\right\vert ^{2}\right) dx=0.  \label{complex}
\end{equation}%
The substitution of the zero-order wave function in Eq. (\ref{complex})
predicts the squared amplitude,%
\begin{equation}
\left( A_{0}^{\mathrm{(1D)}}\right) ^{2}=\frac{4\gamma -\Omega \eta }{2\sqrt{%
2}\Gamma }.  \label{A0^2}
\end{equation}%
The norm (\ref{N}) of the solution with amplitude (\ref{A0^2}) is%
\begin{equation}
N=\frac{\sqrt{\pi }\left( 4\gamma -\Omega \eta \right) }{2\sqrt{2\Omega }%
\Gamma }.  \label{N_sigma=0}
\end{equation}%
In particular, Eq. (\ref{N_sigma=0}) predicts that nonzero states exist at $%
\gamma >\gamma _{\mathrm{thr}}=\Omega \eta /4$. For small $\eta $, this
result agrees with the exact threshold given by Eq. (\ref{eta_thr-exact}).

A similar prediction can be elaborated for 2D vortex modes produced by Eq. (%
\ref{u2D}) with $\sigma =0$ and $\Gamma >0$. In this case, the zero-order
wave function is $u_{0}^{\mathrm{(2D)}}(r)=A_{0}^{{\Large (\mathrm{2D})}%
}r^{S}\exp \left( -\Omega r^{2}/2\right) $, cf. Eq. (\ref{lin2D}), while the
2D power-balance condition is%
\begin{equation}
\int_{0}^{\infty }\left[ \gamma |u|^{2}-\Gamma |u|^{4}-\frac{1}{2}\eta
\left( \left\vert \frac{du}{dr}\right\vert ^{2}+\frac{S^{2}}{r^{2}}%
|u|\right) \right] rdr=0.  \label{balance}
\end{equation}%
The substitution of $u_{0}^{\mathrm{(2D)}}(r)$ in Eq. (\ref{balance}) yields
the result:%
\begin{equation}
\left( A_{0}^{\mathrm{(2D)}}\right) ^{2}=2^{1+2S}\frac{S!}{\left( 2S\right) !%
}\frac{\Omega ^{S}}{\Gamma }\left( \gamma -\frac{1+S}{2}\Omega \eta \right) ,
\label{A0^2-2D}
\end{equation}%
which corresponds to the 2D integral power (\ref{N2D})%
\begin{equation}
N=2^{1+2S}\frac{\pi S!}{(2S)!}\left( \Omega \Gamma \right) ^{-1}\left(
\gamma -\frac{1+S}{2}\Omega \eta \right) ,  \label{N-2D}
\end{equation}%
cf. Eq. (\ref{N_sigma=0}). The threshold condition, following from Eq. (\ref%
{N-2D}), $\gamma >\gamma _{\mathrm{thr}}\left( 1+S\right) /2$, in the limit
of small $\eta $ is consistent with the exact 2D threshold condition (\ref%
{eta_thr_2D}).

The analytical results, both exact and analytical ones, are compared to
their numerical counterparts below.

\section{Basic results: the 1D setting}

Numerical solutions of stationary equations (\ref{u}) and (\ref{u2D}) were
constructed by means of the squared-operator iterative method \cite{SOM}.
The evolution governed by Eqs. (\ref{1}) and (\ref{2D}) was simulated using
the fourth-order split-step Fourier-transform algorithm. The integration
domain with dimensions $-8<x,y<+8$ was sufficient to accommodate all
confined modes investigated herein. In most cases, the numerical grid of
size $128\times 128$ was sufficient to produce the modes with necessary
accuracy. The stability and instability of all stationary modes, predicted
by the numerical solution of linearized equations (\ref{pert}), was
accurately corroborated by direct simulations. As said above, control
parameters of the model are $\gamma =\Gamma $ [see Eq. (\ref{gg})], $\eta $,
and $\Omega ^{2}$, as well as the discrete nonlinearity coefficient, $\sigma
=-1,0,+1$ in Eq. (\ref{sigma}).

\begin{figure}[tbph]
\centering\includegraphics[width=3in]{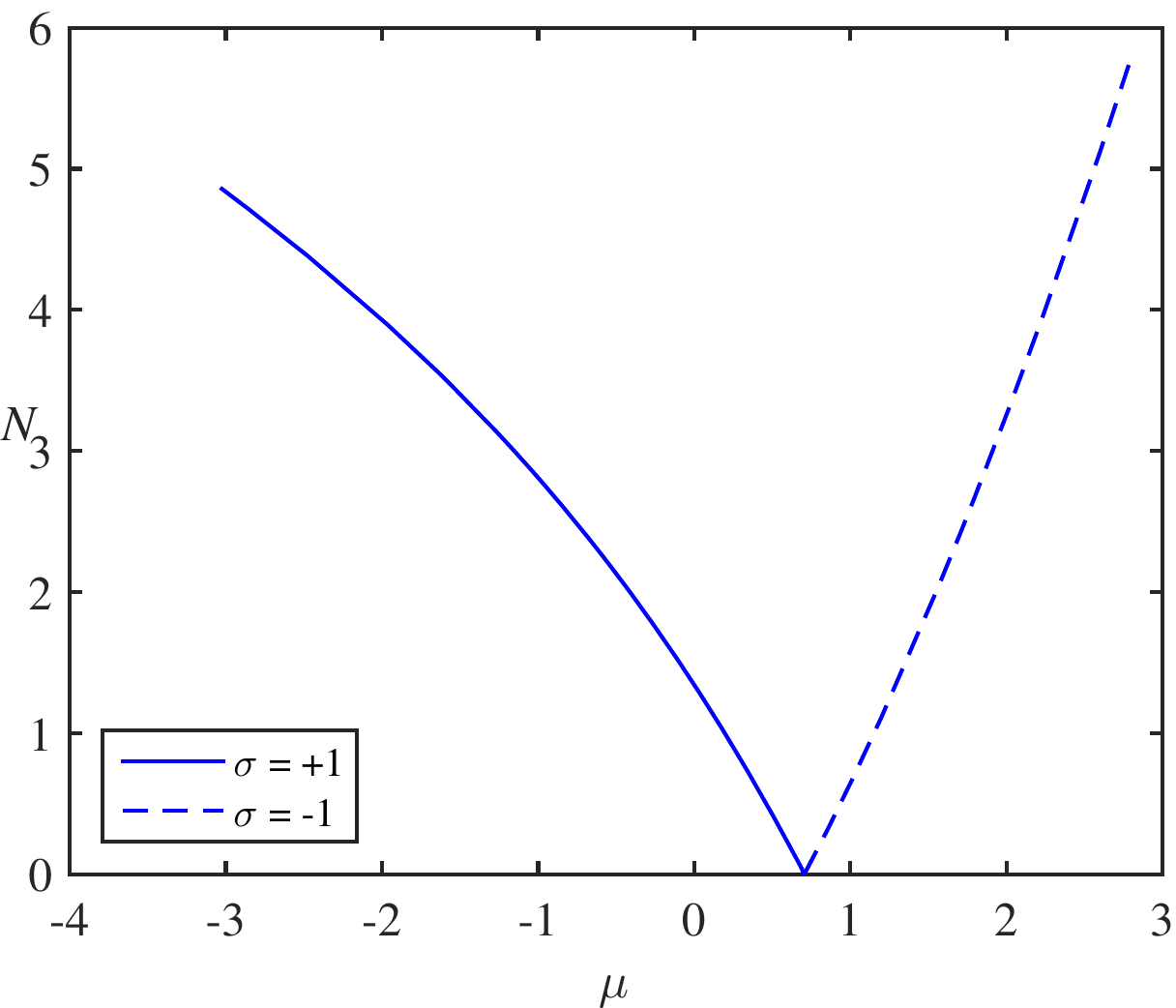}
\caption{The integral power (norm) of 1D ground states in the conservative
limit, $\protect\gamma =\Gamma =\protect\eta =0$, vs. the chemical
potential, for the self-focusing ($\protect\sigma =+1$) and defocusing ($%
\protect\sigma =-1$) signs of the cubic term, with the strength of the HO
potential $\Omega ^{2}=2$.}
\label{fig1}
\end{figure}

First, in the absence of losses, $\gamma =\Gamma =\eta =0$, it is easy to
construct a family of real GS solutions of Eq. (\ref{u}) for either sign of $%
\sigma $, which correspond to stable solutions of Eq. (\ref{1}). For $\sigma
=+1$ and $-1$ (the self-focusing/defocusing cubic term), the shape of the GS
is close, respectively, to that of the usual bright soliton placed at the
bottom of the HO potential trap, or to the GS predicted by the Thomas-Fermi
(TF) approximation \cite{TF}. In both cases, the families of the GSs are
characterized by dependences of their integral power (\ref{N}) on chemical
potential $\mu $, as shown in Fig. \ref{fig1}. Naturally, in the limit of $%
N\rightarrow 0$ the branches originate from value $\mu =\Omega /2$, which
corresponds to the GS of the HO potential, while in the limit of $%
N\rightarrow \infty $ the dependences are determined, respectively, by the
free-space soliton solution and the TF approximation : $N_{\mathrm{sol}}(\mu
)=2\sqrt{-2\mu }$; $N_{\mathrm{TF}}(\mu )=2(3\Omega )^{-1}\left( 2\mu
\right) ^{3/2}$.

\begin{figure}[tbph]
\centering\includegraphics[width=3in]{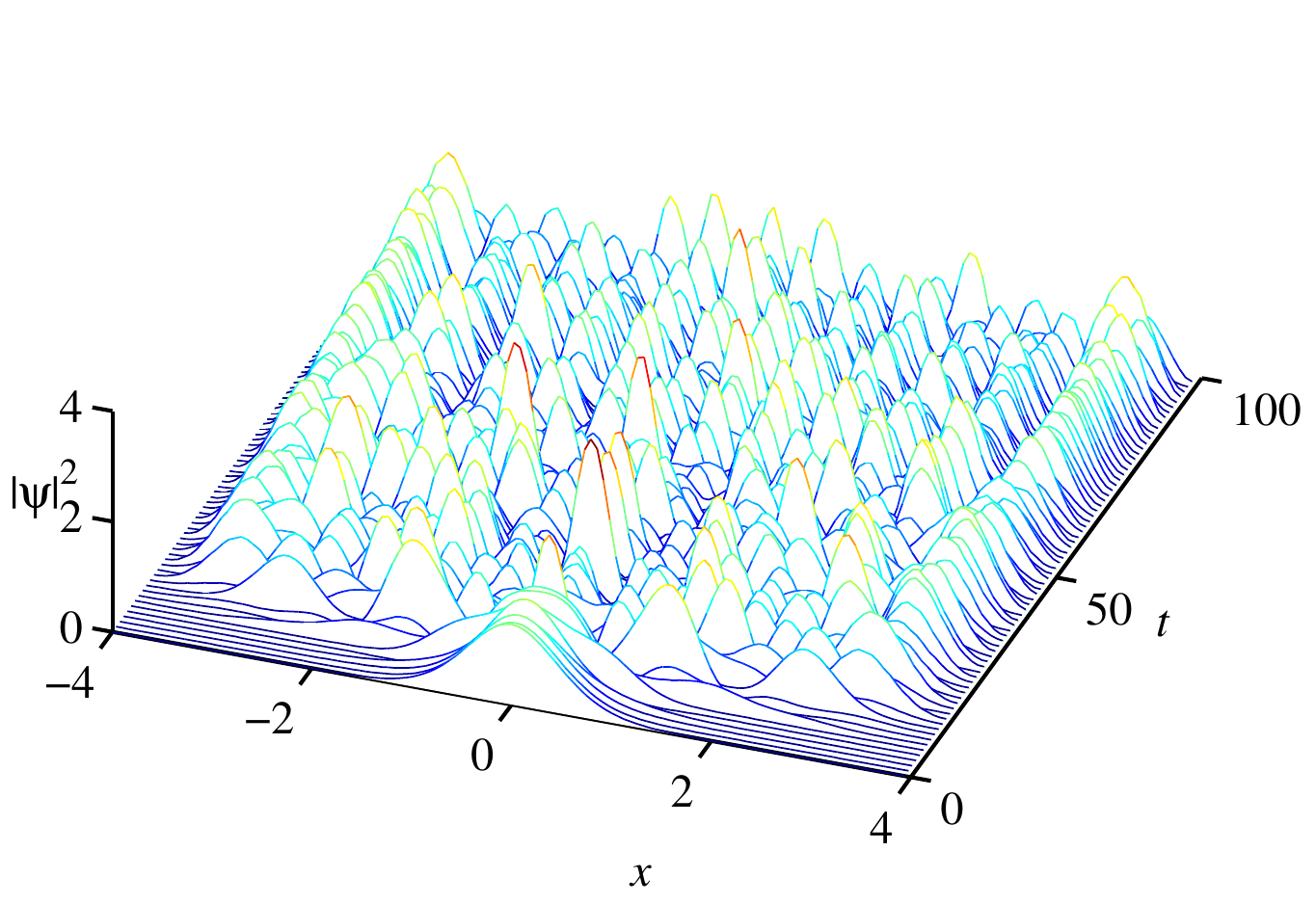}
\caption{The unstable evolution of a numerically constructed stationary
trapped 1D mode for $\protect\gamma =\Gamma =1.0$, $\protect\eta =0$ (no
diffusivity), $\Omega ^{2}=2$, and $\protect\sigma =+1$ (the self-focusing
cubic term).}
\label{fig2}
\end{figure}

Proceeding to the 1D model with $\gamma =\Gamma =1$, we note, first, that
all the stationary modes are unstable in the absence of the effective
diffusion, $\eta =0$, similar to what was reported in Ref. \cite{Sakaguchi}.
This instability transforms static modes into quasi-turbulent states which
fill the entire space admitted by the HO trapping potential, as shown in
Fig. \ref{fig2}.

\begin{figure}[tbph]
\centering\includegraphics[width=5in]{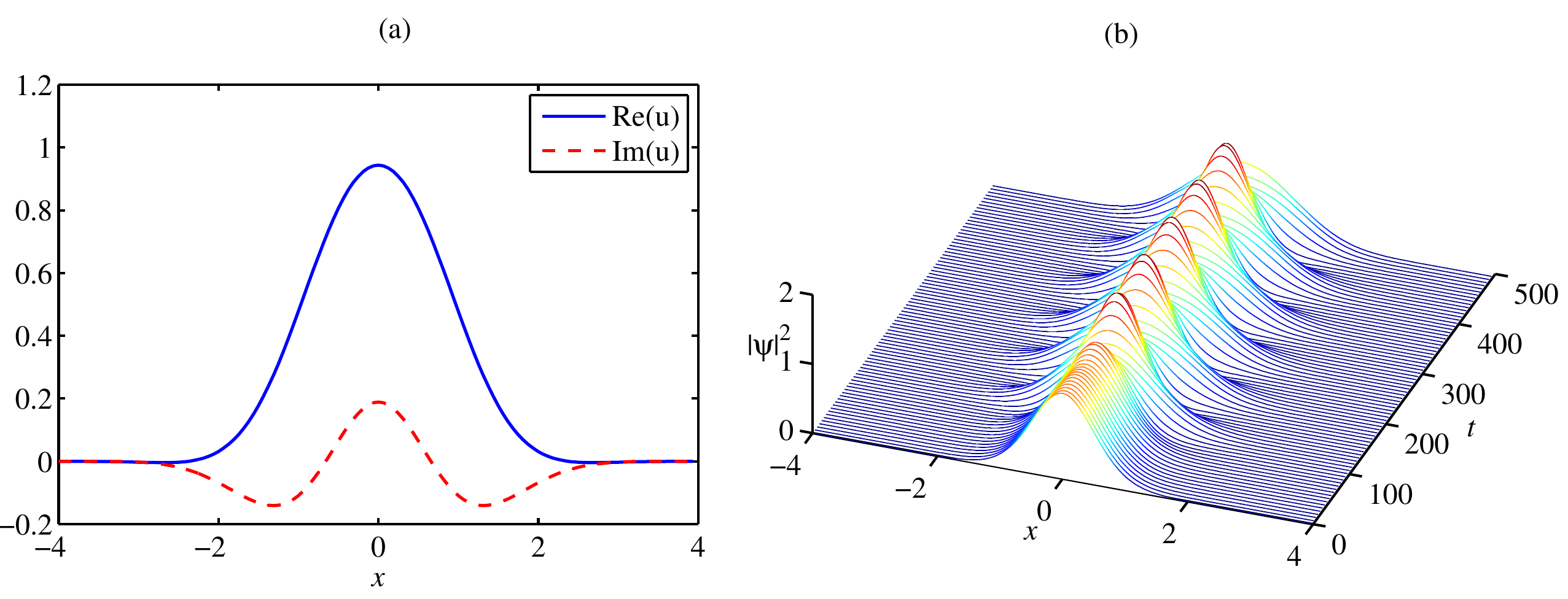}
\caption{Typical examples of robust dynamical regimes in the 1D model. (a)
Real and imaginary parts of a typical stable stationary ground-state mode,
found at $\protect\eta =0.8$, $\Omega ^{2}=2$. The chemical potential and
integral power (norm) of this mode are $\protect\mu =0.1714$ and $N=1.4120$.
(b) A persistent breather spontaneously generated by an unstable stationary
mode, at $\protect\eta =0.3$. In both cases, $\protect\gamma =\Gamma =1$, $%
\Omega ^{2}=2$, and $\protect\sigma =+1$ (self-focusing).}
\label{fig3}
\end{figure}

\begin{figure}[tbph]
\centering\includegraphics[width=5in]{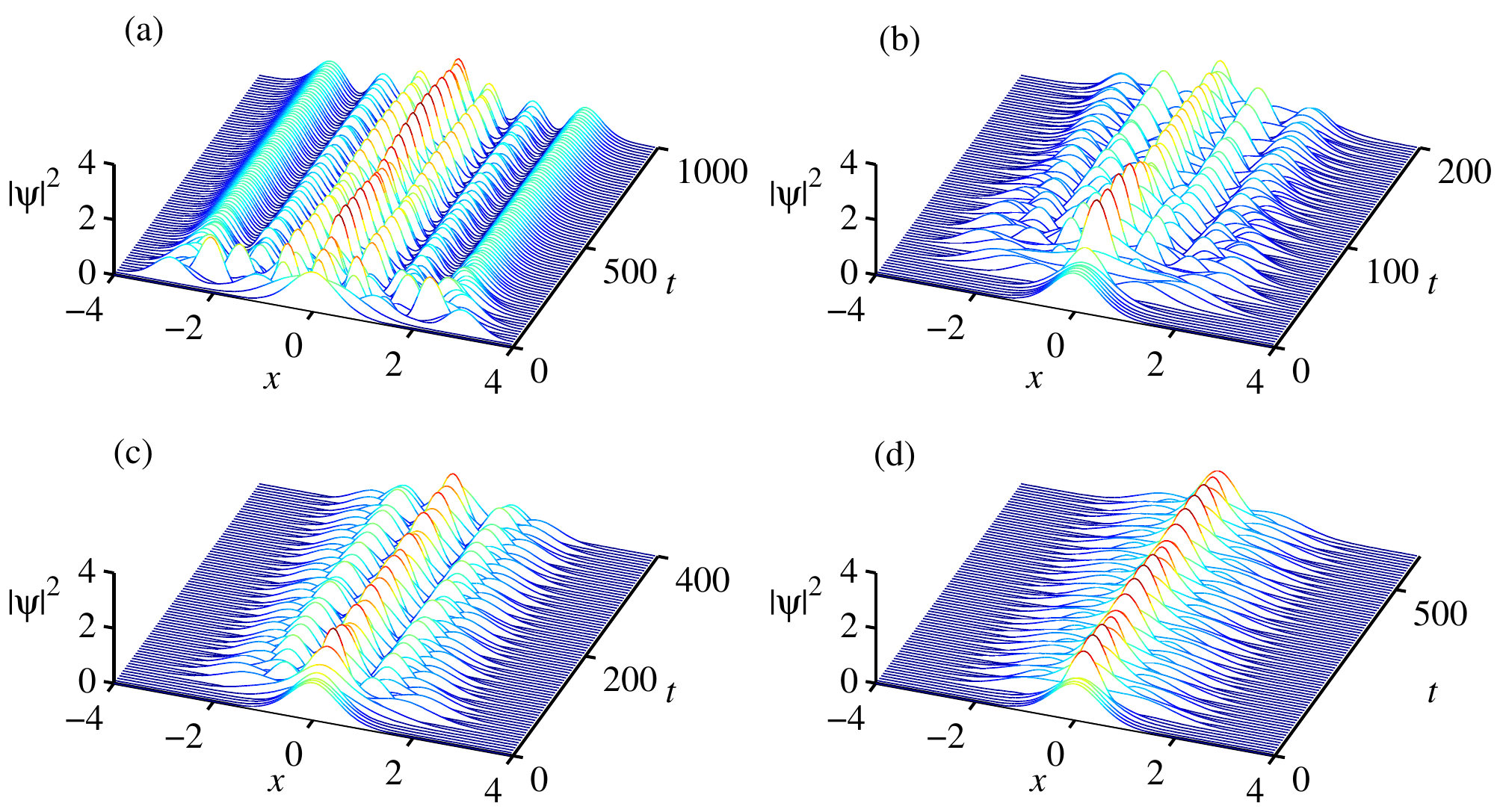}
\caption{Robust quasi-regular oscillatory states with seven (a), five (b),
three (d), and single (d) peaks, developing from unstable stationary modes
at $\protect\eta =0.05,0.10,0.15$, and $0.2$, respectively. Other parameters
are the same as in Fig. \protect\ref{fig3}.}
\label{fig4}
\end{figure}

Robust stationary and dynamical states are found in the presence of $\eta >0$%
: stable stationary GS modes existing at relatively large values of $\eta $
[Fig. \ref{fig3}(a)]; persistent breathers, spontaneously emerging from some
unstable stationary modes at smaller $\eta $ [Fig. \ref{fig3}(b)]; and
quasi-regular multi-peak periodically oscillating states, developing from
unstable stationary modes at small $\eta $ [Fig. \ref{fig4}]. In the limit
of $\eta \rightarrow 0$, the latter pattern goes over into the
quasi-turbulent one displayed in Fig. \ref{fig2}. The increase of $\eta $
naturally leads to simplification of the quasi-regular patterns via the
decrease of the number of peaks, from seven in Fig. \ref{fig4}(a) to five in %
\ref{fig4}(b), to three \ref{fig4}(c) and, eventually, to one in Fig. \ref%
{fig4}(d). The latter dynamical mode is close to a regular breather, cf.
Fig. \ref{fig3}(b).

\begin{figure}[tbph]
\centering\includegraphics[width=3in]{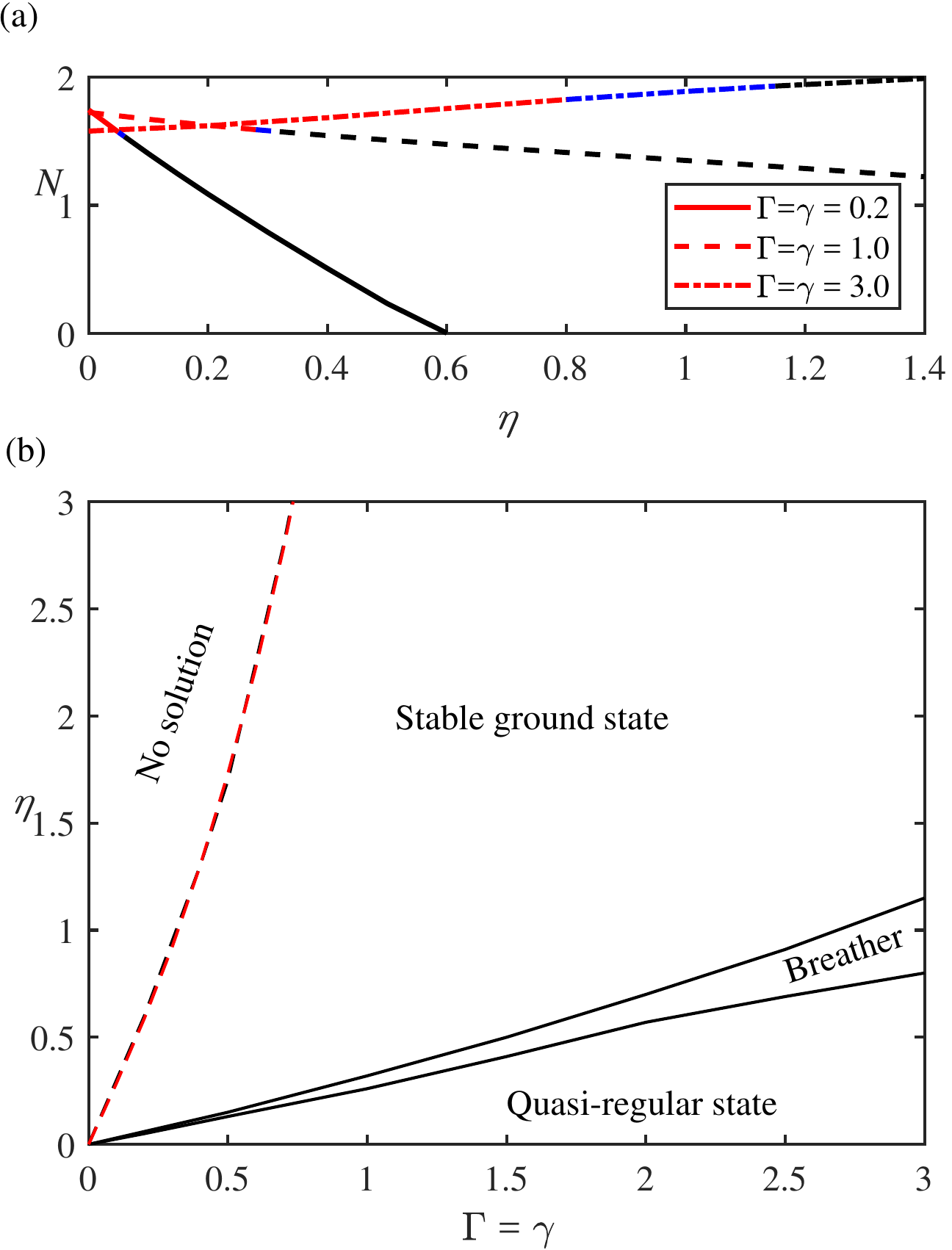}
\caption{(a) The integral power (norm) of the numerically generated
stationary 1D modes vs. diffusivity $\protect\eta $, for $\protect\sigma =+1$
(the self-focusing cubic term), $\Omega ^{2}=2$, and $\protect\gamma =\Gamma
=0.2,\,1.0$, and $3.0$ (solid, dashed, and dotted lines, respectively). The
stability is identified by colors: red corresponds to unstable stationary
modes generating quasi-regular states; blue corresponds to unstable modes
developing into breathers; and black represents stable stationary modes. (b)
The stability map for different robust stationary and dynamical states,
labeled in the plane of ($\protect\gamma =\Gamma $,$\protect\eta $), for $%
\Omega ^{2}=2$ and $\protect\sigma =+1$. The dashed line is the existence
boundary for the stationary states, as predicted by Eq. (\protect\ref%
{eta_thr-exact}). It completely coincides with the numerically found
counterpart.}
\label{fig5}
\end{figure}

The results obtained in the case of self-focusing, $\sigma =+1$, are
summarized in Fig. \ref{fig5}, where panel (a) shows the numerically found
dependence of the integral power, $N$ on the crucially important control
parameter, $\eta $, and panel (b) presents the findings in the form of
stability map displayed in the plane of $\gamma =\Gamma $ and $\eta $. In
particular, we stress that a numerically found existence boundary for stable
GSs, shown by the black dashed line in (b), \emph{precisely} coincides with
the analytical prediction given by Eq. (\ref{eta_thr-exact}), which is shown
by the red dashed line. Further, we observe that vast areas populated by
stable GSs and persistent quasi-regular states are separated by a narrow
sliver supporting stable breathers. We also stress that no case of
bistability was produced by the systematic numerical analysis.

\begin{figure}[tbph]
\centering\includegraphics[width=3in]{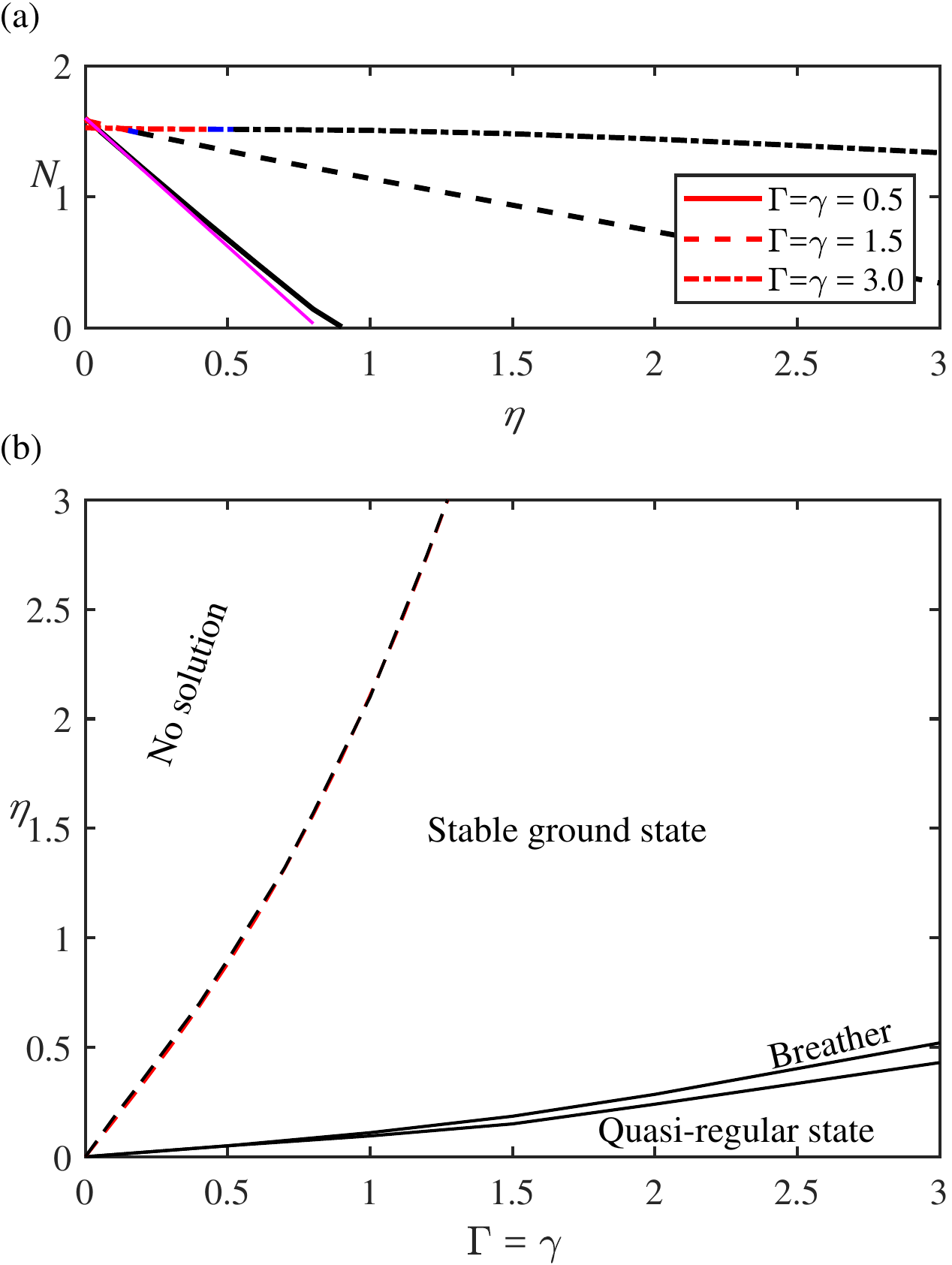}
\caption{The same as in Fig. \protect\ref{fig5}, but for the 1D model with $%
\protect\sigma =0$ and $\Omega ^{2}=6$. The magenta solid line in (a)
additionally displays the analytical approximation (\protect\ref{N_sigma=0})
for $\protect\gamma =\Gamma =0.5$.}
\label{fig6}
\end{figure}

In the 1D model with the self-focusing cubic term, $\sigma =+1$, the results
obtained for other values of the HO potential strength, $\Omega ^{2}$, are
quite similar to those displayed above for $\Omega ^{2}=1$. Further, the
results also remain similar for the defocusing sign of the cubic term, $%
\sigma =-1$, as well as in the 1D model with $\sigma =0$, where the
nonlinearity is represented solely by the cubic loss, $\Gamma >0$. These
conclusions are illustrated by Fig. \ref{fig6}, which summarizes the results
for $\sigma =0$ and $\Omega ^{2}=6$, in the same way as it is done in Fig. %
\ref{fig5} for $\sigma =+1$ and $\Omega ^{2}=2$. In particular, the magenta
line in panel \ref{fig5} demonstrates that the analytical approximation,
elaborated for $\sigma =0$ in the form of Eq. \ref{N_sigma=0}), is an
accurate one. Overlapping black and red dashed curves display, severally,
the numerically found and analytically predicted [see Eq. (\ref%
{eta_thr-exact})] existence boundaries for the stationary GS modes.

\section{Basic results: the 2D setting}

\begin{figure}[tbph]
\centering\includegraphics[width=3in]{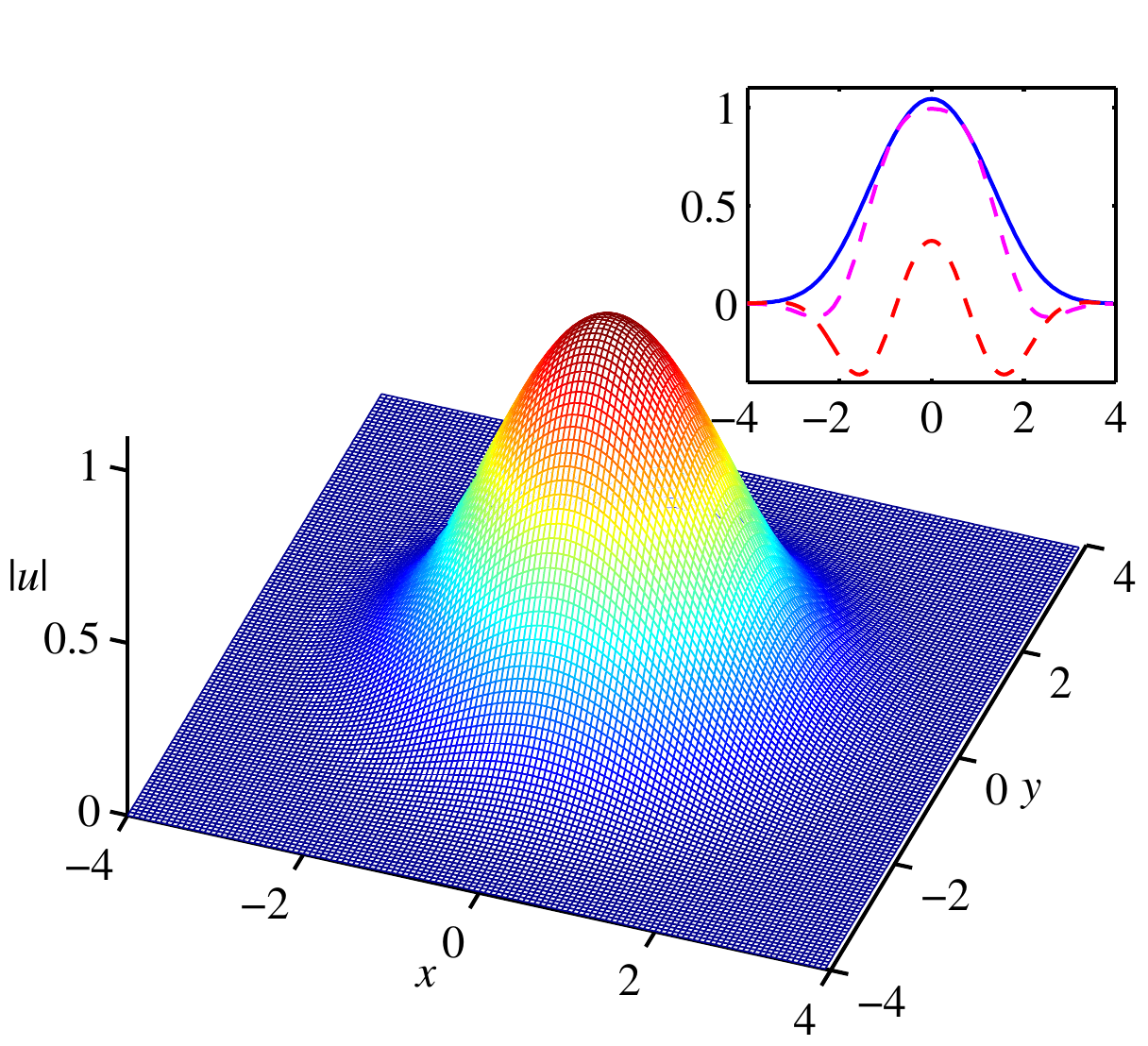}
\caption{A stable 2D GS, numerically found for $\protect\sigma =0$, $\Gamma =%
\protect\gamma =3.0$, $\protect\eta =1.5$, and $\Omega ^{2}=2$. The inserted
panel shows radial profiles, along $y=0$, of the amplitude $|u\left(
x\right) |$ (blue), as well as real (magenta) and imaginary (red) parts of $%
u(x)$.}
\label{fig7}
\end{figure}

In the 2D model, the systematic numerical analysis has identified several
types of stationary and dynamical states. First, similar to the 1D case, the
largest stability area is found for stationary GS states, which are trapped
isotropic modes, see an example in Fig. \ref{fig7}. Next, persistent
axisymmetric breathers, developing from unstable stationary states, are
found too, although in a small parameter area (see below). An example of the
breather is displayed in Fig. \ref{fig8}
\begin{figure}[tbph]
\centering\includegraphics[width=3in]{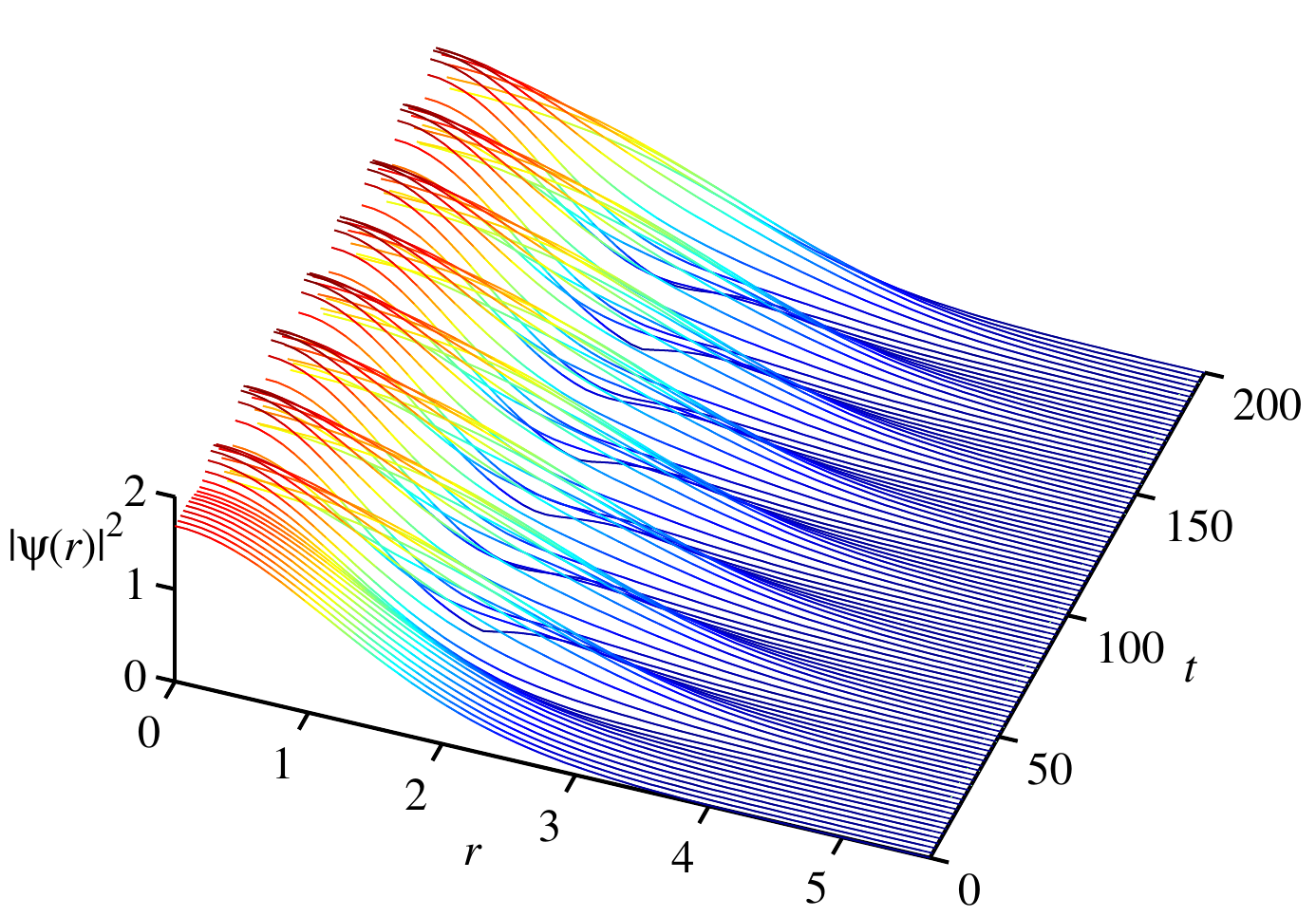}
\caption{A persistent isotropic breather generated by an unstable stationary
mode at $\protect\sigma =+1$, $\Gamma =\protect\gamma =2.3$, $\protect\eta %
=0.6$, and $\Omega ^{2}=2$.}
\label{fig8}
\end{figure}

\begin{figure}[tbph]
\centering\includegraphics[width=5in]{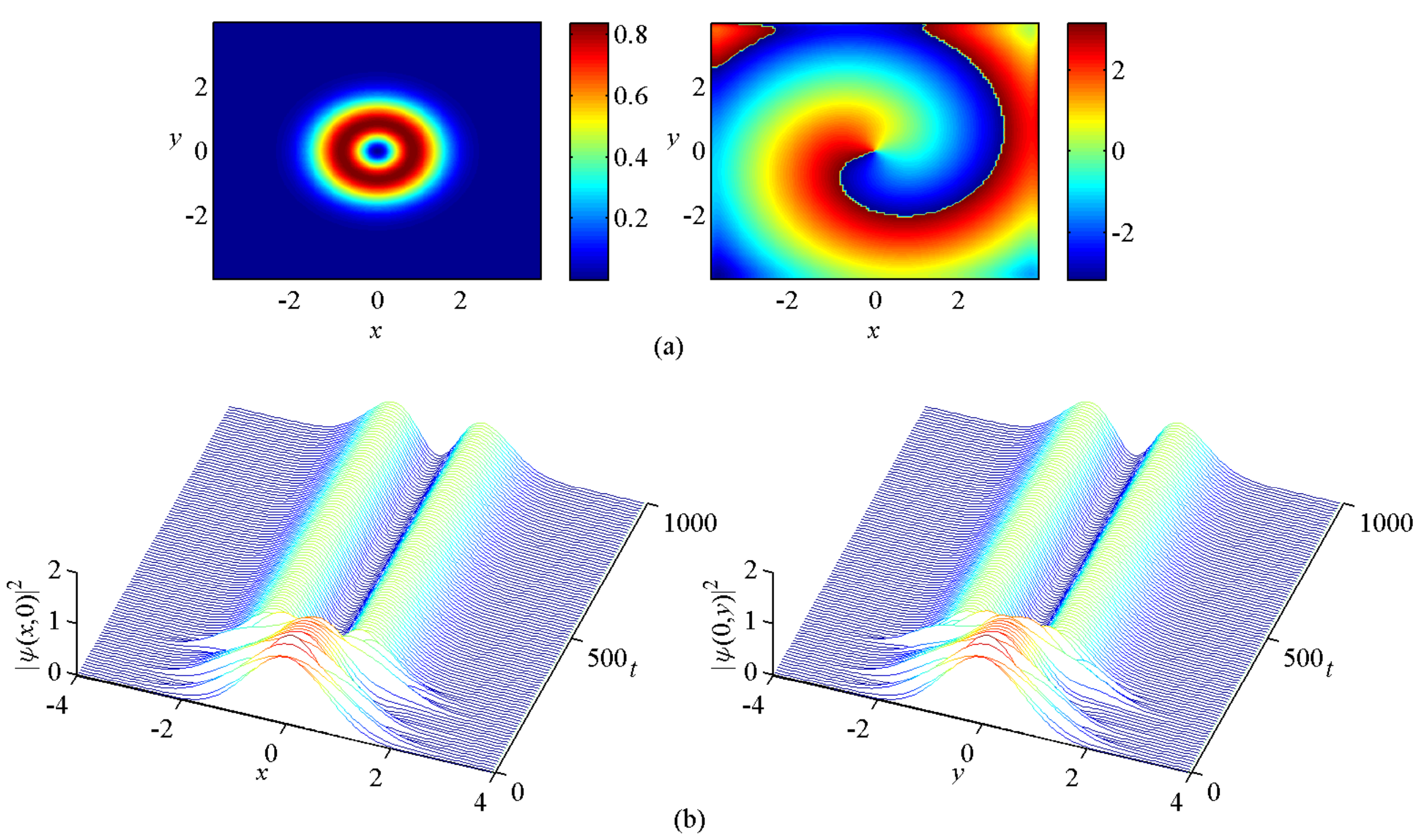}
\caption{(a) The local-intensity [$|u(x,y)|^{2}$] and phase structure of a
stable axisymmetric vortex spontaneously generated by an unstable stationary
mode, at $\protect\sigma =0$, $\Gamma =\protect\gamma =2.5$, $\protect\eta %
=0.6$, and $\Omega ^{2}=2$. (b) The evolution of the local intensity in the $%
x$ and $y$ cross-sections, illustrating the spontaneous transformation of
the unstable stationary mode into the isotropic vortex.}
\label{fig9}
\end{figure}

The decrease of $\eta $ leads, as in the 1D model, to destabilization of the
stationary modes. The evolution of unstable modes generates a variety of
different dynamical states featuring vortex structures, cf. Refs. \cite%
{Skarka1,Skarka2}, where the 2D CGLE with the cubic-quintic nonlinearity and
a radially localized linear gain (but without a trapping potential and
diffusion) also gave rise to a large number of different vortical modes;
stable complexes built of one, two, or three vortices were reported too in
the study of the cubic-quintic CGLE model with the uniform linear loss and
trapping HO potential \cite{Skarka}. Among such structures, we first
identify stable axisymmetric vortices with topological charge $S=1$, see an
example in \ref{fig9}. The spontaneous generation of vorticity in the
dissipative medium is explained by the fact that, originally, it generates a
vortex-antivortex pair, whose antivortex component disappears in the
peripheral area. The axisymmetric vortices have also been produced as
stationary solutions of Eq. (\ref{u2D}), and their stability has been
identified by means of Eq. (\ref{pert2D}).

\begin{figure}[tbph]
\centering\includegraphics[width=5in]{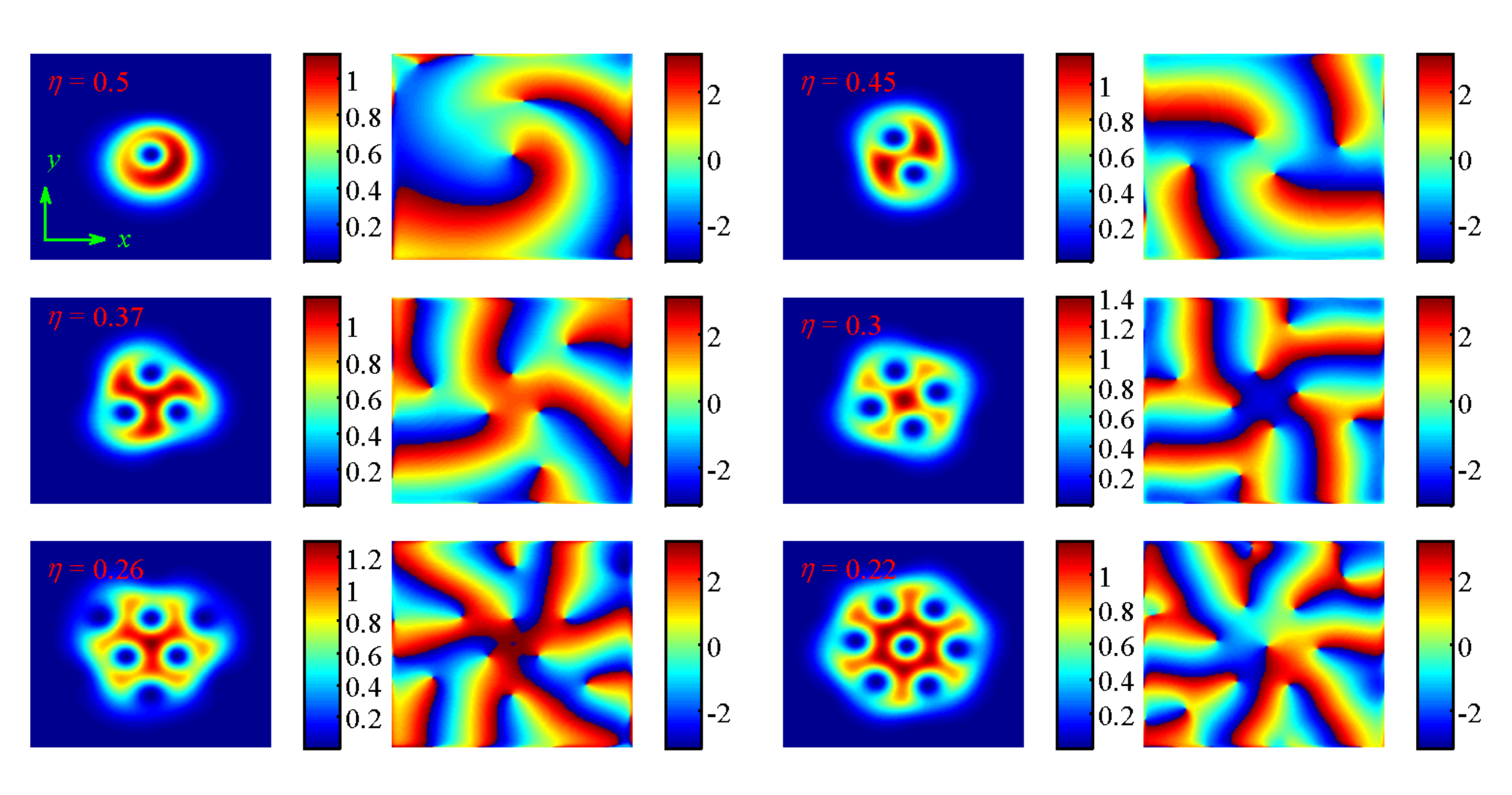}
\caption{A sequence of stably rotating states produced by the simulations of
the 2D model for $\protect\sigma =+1$, $\Gamma =\protect\gamma =2.5$, and $%
\Omega ^{2}=2$, at decreasing values of diffusivity $\protect\eta $ (which
are indicated in the panels): a crescent vortex ($S=1$), and multi-vortex
complexes with $S=2,3,4,6,$ and $7$. At $\protect\eta >0.5$, the model
supports a a stable axisymmetric vortex with $S=1$, while at $\protect\eta %
<0.22$ a transition to vortex turbulence occurs. In all panels, spatial
scales are the same as in Fig. \protect\ref{fig9}(a).}
\label{fig10}
\end{figure}

\begin{figure}[tbph]
\centering\includegraphics[width=5in]{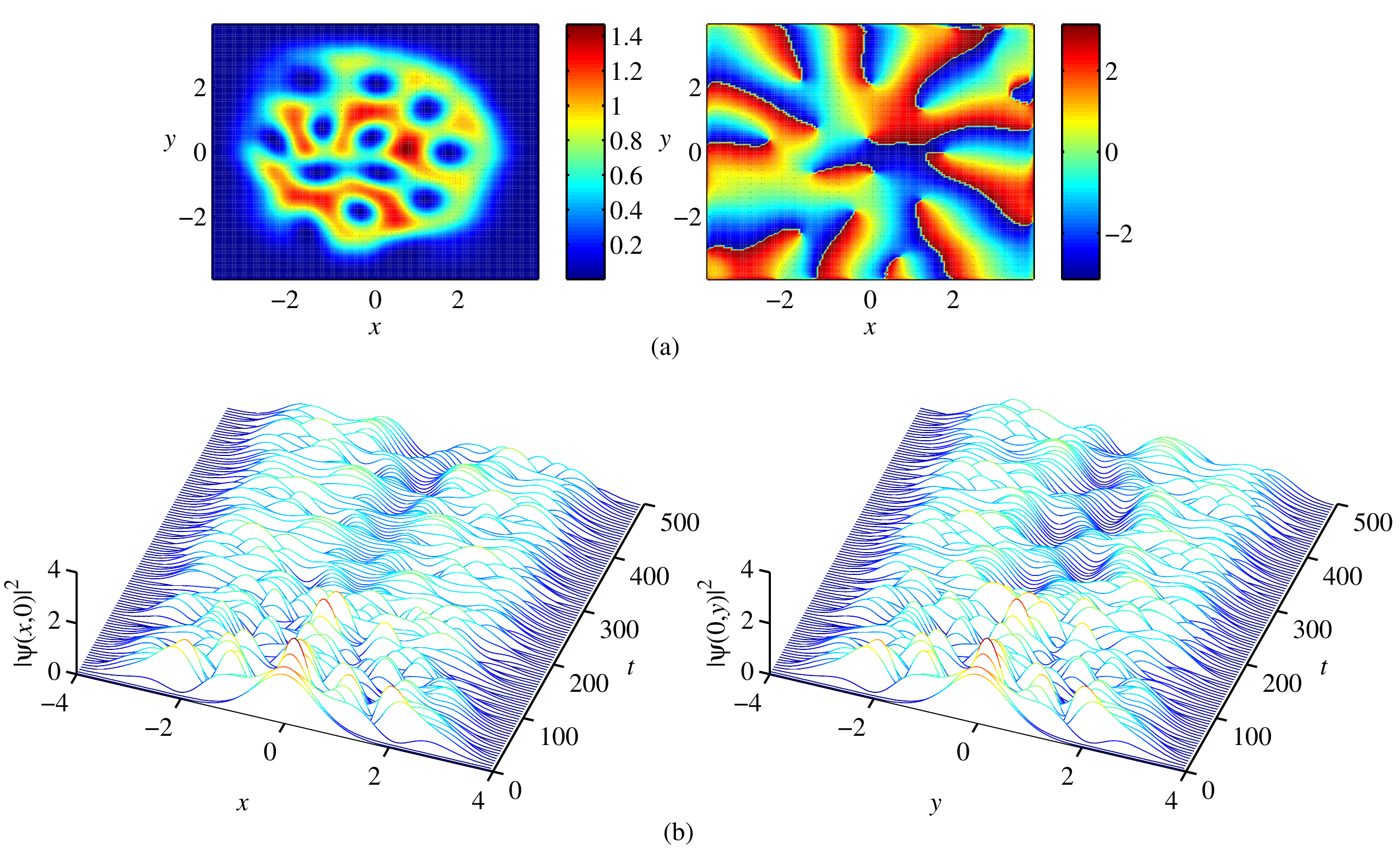}
\caption{An example of the vortex-turbulent state, found for $\protect\sigma %
=0$, $\protect\gamma =\Gamma =2$, $\protect\eta =0.1$, and $\Omega ^{2}=2$.
(a) Snapshots of the local-intensity, $|\protect\psi \left( x,y\right) |^{2}$%
, and phase patterns (left and right panels, respectively), produced, at $%
t=1000$, by simulations initiated with an unstable stationary state. (b) The
respective evolution of the local intensity, shown in the $x$- and $y$-
cross sections.}
\label{fig11}
\end{figure}

Further decrease of $\eta $ causes destabilization of the axisymmetric
vortices and replaces them, depending on values of other parameters
(especially, $\sigma $), either by rotating azimuthally nonuniform vortices
with a crescent shape and $S=1$, or by rotating two-vortex complexes, with $%
S=2$ (see examples in Fig. \ref{fig10}), or by a state which may be
categorized as vortex turbulence, see Fig. \ref{fig11}. In particular, the
crescents emerge only in the case of the self-focusing cubic nonlinearity, $%
\sigma =+1$. Subsequent decrease of $\eta $ reveals transition to more
complex rotating sets of vortices, which feature the net topological charge
up to $S=8$, see a set of typical examples in Fig. \ref{fig10} (in this
particular figure, the largest topological charge of a stable multi-vortex
complex is $S_{\max }=7$; a peculiarity of the latter bound state, displayed
in the bottom panels of Fig. \ref{fig10}, is that it has one vortex placed
at the center, surrounded by a rotating necklace of six satellites).
Eventually, at very small values of $\eta $, the 2D model gives rise to the
above-mentioned vortex-turbulent state, an example of which is presented in
Fig. \ref{fig11}.

Numerically measured characteristics of the rotating complexes displayed in
Fig. \ref{fig10}, \textit{viz}., the values of their radius, defined as the
distance of maxima of the local intensity from the rotation pivot, and the
angular velocity of the rotation, are collected in Table \ref{table}.
Naturally, the radius tends to grow with the increase of the total number of
individual vortices, $S$. The value of radius $\rho $\ of the ring-shaped
crescent vortex (as well as the radius of the isotropic vortex, when it is
stable) can be easily explained as one at which the density of the
eigenstate with $S=1$\ in the linear Schr\"{o}dinger equation with the HO
potential attains the maximum, which is $\rho _{0}=\Omega ^{-1/4}$. Indeed, $%
\rho _{0}(\Omega =2)\approx \allowbreak 0.8409$\ is almost identical to the
numerically determined radius for $S=1$\ in Table \ref{table}. As concerns
the angular velocity, $\omega $, of the rotating crescent, we note that, for
a narrow ring of radius $\rho $, the linear Schr\"{o}dinger equation
produces an elementary result, $\omega =S/\rho ^{2}$. In reality, the
crescent, displayed in the top row of Fig. \ref{fig10}, is not quite narrow,
and the comparison of this formula with the respective angular velocity
given in Table \ref{table} suggests that $\omega $\ is determined by the
inner layer of the crescent, which has $\rho \simeq 0.5$. It is relevant to
note that the rotation velocity of the crescent-shaped vortex with $S=1$ is
essentially higher than in all the complexes with $S\geq 2$.
\begin{table}[tbp]
\caption{Characteristics of rotating complexes built of $S$ individual
vortices shown in Fig. \protect\ref{fig10}, i.e., with $\protect\sigma =+1$,
$\Gamma =\protect\gamma =2.5$, and $\Omega ^{2}=2$ ($S=1$ corresponds to the
rotating crescent-shaped vortex).}
\label{table}\centering
\begin{tabular}{c|c|c}
Total vorticity & the effective radius & $\omega$ \\ \hline
$S=1$ ($\eta =0.50$) & $0.8359$ & $3.927$ \\
$S=2$ ($\eta =0.45$) & $0.7500$ & $1.6526$ \\
$S=3$ ($\eta =0.37$) & $1.0000$ & $1.5932$ \\
$S=4$ ($\eta =0.30$) & $1.1250$ & $1.5150$ \\
$S=6$ ($\eta =0.26$) & $1.4688$ & $1.4956$ \\
$S=7$ ($\eta =0.22$) & $1.7500$ & $1.4952$%
\end{tabular}%
\end{table}

\begin{figure}[tbph]
\centering\includegraphics[width=3in]{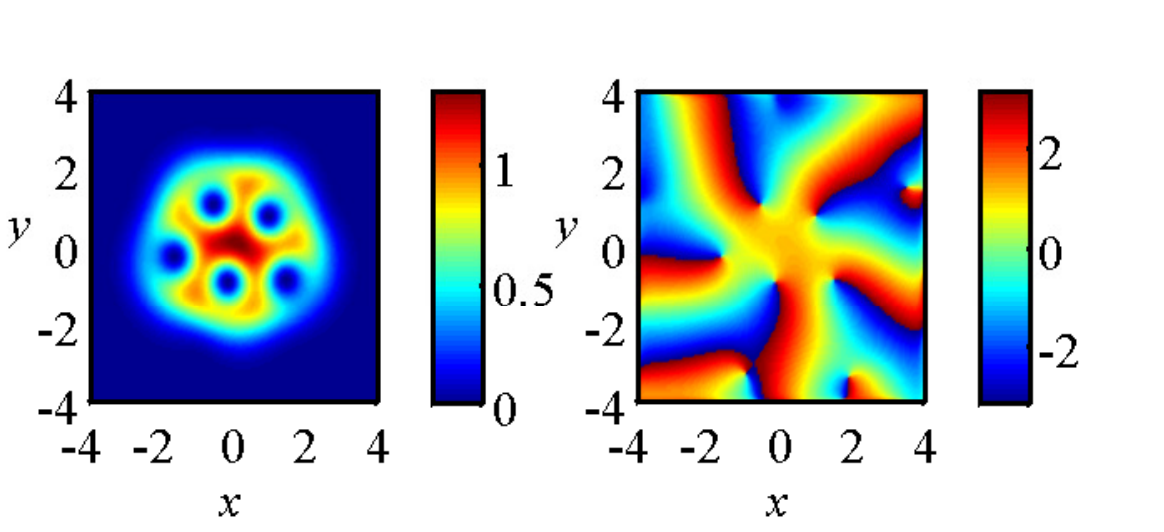}
\caption{The local-intensity and phase structure of a stably rotating $5$%
-vortex bound state found for $\protect\sigma =-1$, $\Gamma =\protect\gamma %
=2.0$, $\protect\eta =0.23$, and $\Omega ^{2}=2$.}
\label{fig12}
\end{figure}

Note that the sequence of multi-vortex complexes, displayed in Fig. \ref%
{fig10} for $\sigma =+1$ (the self-focusing sign of the nonlinearity), does
not include a bound state of five vortices. Actually, such a stable state
can be found too, but for $\sigma =-1$, as shown in Fig. \ref{fig12}.

\begin{figure}[tbph]
\centering\includegraphics[width=4.5in]{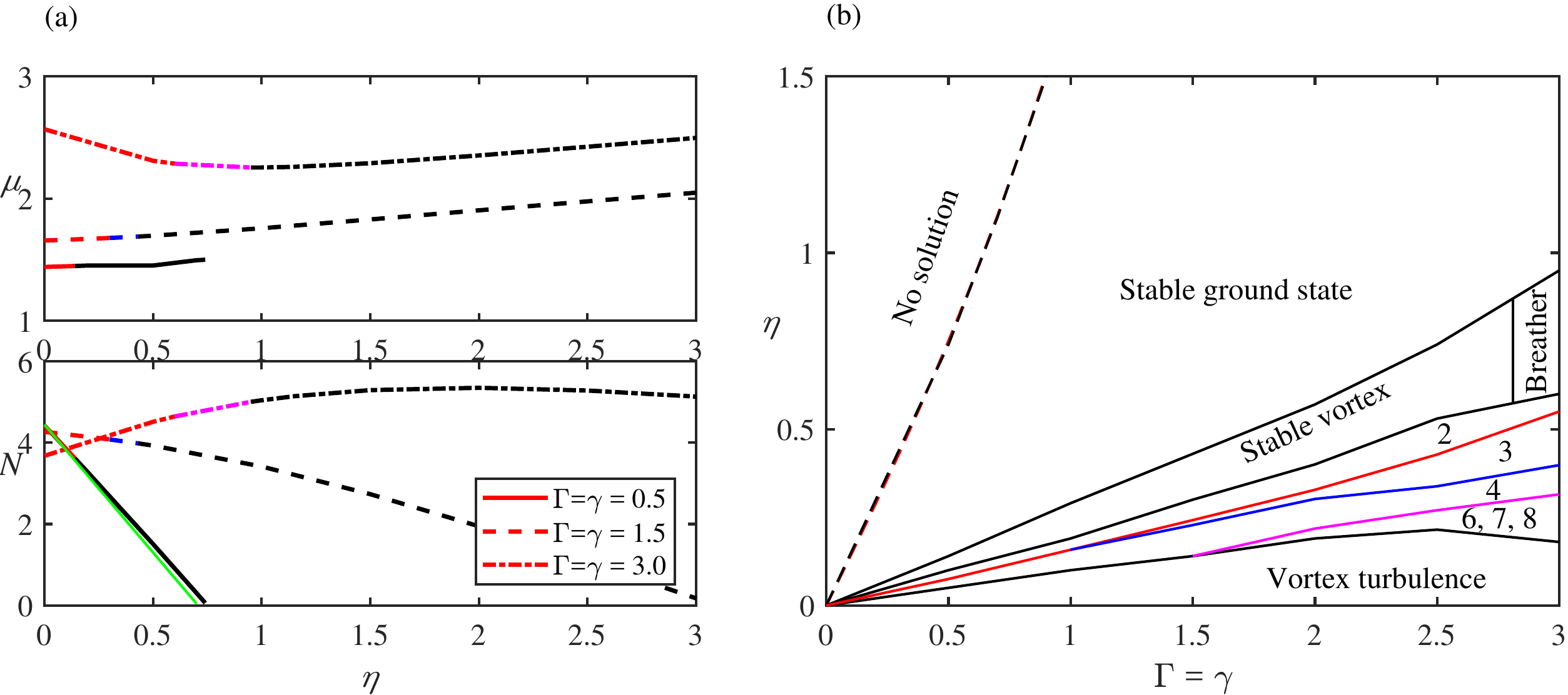}
\caption{(a) The integral power (norm) of numerically found 2D stationary
modes, given by Eq. (\protect\ref{N2D}), vs. diffusivity $\protect\eta $ for
$\protect\sigma =0$, $\Omega ^{2}=2$, and $\protect\gamma =\Gamma =0.5$, $%
1.5 $, and $3.0$ (solid, dashed, and dotted lines, respectively). The
stability is identified by colors: red, blue, and magenta imply, severally,
the transformation into vortex tubulence, stable axisymmetric vortices, and
breathers, while black curves represent families of stable stationary modes.
The green solid line displays the analytical approximation (\protect\ref%
{N-2D}) for $\protect\gamma =\Gamma =0.5$. (b) Stability borders in the
plane of $\left( \protect\gamma =\Gamma ,\protect\eta \right) $ for $\protect%
\sigma =0$ and $\Omega ^{2}=1$. The analytical existence boundary, $\protect%
\eta _{\mathrm{thr}}^{(\mathrm{2D})}$, predicted by Eq. (\protect\ref%
{eta_thr_2D}), and its numerically identified counterpart are shown,
respectively, by red and black dashed lines, which completely overlap.
Digits in subregions represent the number of individual vortices in rotating
complexes populating these subregions. In particular, $1$ implies the
presence of the single ($S=1$) rotating crescent-shaped vortex, different
from the axisymmetric vortex in the area of \textquotedblleft Stable
vortex". }
\label{fig13}
\end{figure}

\begin{figure}[tbph]
\centering\includegraphics[width=4.5in]{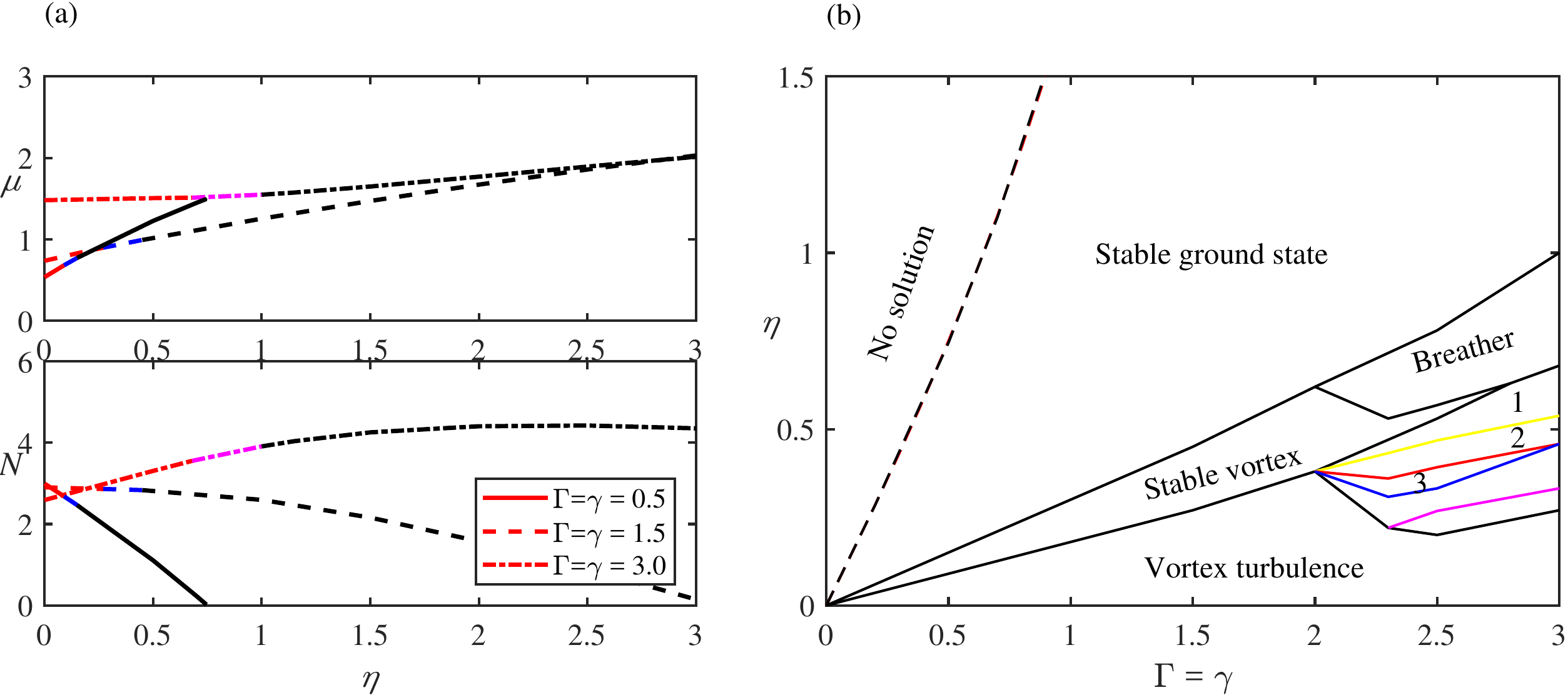}
\caption{The same as in Fig. \protect\ref{fig13}, but for $\protect\sigma %
=+1 $.}
\label{fig14}
\end{figure}

Results produced by systematic numerical analysis of the 2D model are
summarized in Figs. \ref{fig13} and \ref{fig14} for $\sigma =0$ and $\sigma
=+1$, respectively, fixing $\Omega ^{2}=2$. For other values of $\Omega ^{2}$%
, the plots are quite similar to the ones displayed here. Furthermore, for $%
\sigma =-1$, the plots are also similar to what is shown in Fig. \ref{fig13}%
(b) for $\sigma =0$. We stress that the analytical existence boundary for
the stationary mode, predicted by Eq. (\ref{eta_thr_2D}), is completely
identical to its numerically found counterpart. Comparing these stability
maps with their counterparts reported above for the 1D model, cf. Figs. \ref%
{fig6} for $\sigma =0$ and \ref{fig5} for $\sigma =+1$, we conclude that the
axisymmetric vortices with $S=1$ and the vortex-turbulent states play,
roughly speaking, the same role as, respectively, breathers and
quasi-regular patterns in 1D, while the multi-vortex bound states do not
have their 1D counterparts.

\section{Conclusion}

We have demonstrated a new mechanism for the stabilization of confined modes
in the 1D and 2D CGLEs (complex Ginzburg-Landau equations) with the
cubic-only nonlinearity and the spatially uniform linear gain. Namely, we
have found that the background instability is suppressed by the combination
of the HO (harmonic-oscillator) trapping potential and effective diffusion.
Analytical solutions for the linearized version of the CGLE provide
existence boundaries for trapped modes, which are exactly confirmed by
numerical results. Our numerical analysis has produced a set of stable
stationary modes, persistent breathers, and more complex patterns, \textit{%
viz}., quasi-regular multi-peak structures in 1D, and vortices (axisymmetric
and rotating deformed crescent-shaped ones), and rotating multi-vortex
complexes, with the net topological charge up to $S=8$, in 2D. The models
addressed in this work apply to laser cavities and semiconductor
microcavities with exciton-polariton condensates.


\section*{Funding}

Thailand Research Fund (grant BRG6080017); The Royal Society (grant IE
160465); Israel Science Foundation (grant 1286/17); ITMO University Visiting
Professorship (Government of Russia Grant 074-U01); H2020 (691011, Soliring).

\end{document}